\documentstyle[psfig]{mn}
\def\cm{{\rm\thinspace cm}}

\def\erg{{\rm\thinspace erg}}

\def\Mpc{{\rm\thinspace Mpc}}
\def\Msun{\hbox{$\rm\thinspace M_{\odot}$}}

\def\s{{\rm\thinspace s}}
\def\yr{{\rm\thinspace yr}}

\def\Hz{{\rm\thinspace Hz}}

\def\ergpcmsqps{\hbox{$\erg\cm^{-2}\s^{-1}\,$}}

\def\pcmsq{\hbox{$\cm^{-2}\,$}}

\title[Changing X-ray period of IRAS~18325--5926]
{The changing X-ray period of the Seyfert galaxy IRAS18325--5926} 
\author[A.C. Fabian et al]
{\parbox[]{6.5in} { A.C. Fabian$^1$, J.C. Lee$^1$,
W.N. Brandt$^2$, K. Iwasawa$^1$, K.Jahoda$^3$ and C.S.~Reynolds$^4$}\\
\\
$^1$ Institute of Astronomy, Madingley Road, Cambridge CB3 0HA\\
$^2$ Department of Astronomy and Astrophysics, The Pennsylvania State
University, University Park PA 16802, USA\\
$^3$ Laboratory for High Energy Astrophysics, Goddard Space Flight Center,
Greenbelt MD 20771, USA\\
$^4$ JILA, University of Colorado, Campus Box 440, Boulder CO 80309-0440, USA}
\date{}

\begin{document}

\maketitle

\begin{abstract}
We report on two observations of the Seyfert galaxy IRAS18325-5926 made in
1997 December and 1998 February with the Rossi X-ray Timing Explorer (RXTE).
We find evidence for periodicities in the resulting X-ray lightcurves which
are shorter than the 58~ks period found from data of the source taken in
1997 March with the imaging satellite ASCA. It is therefore likely that
IRAS18325-5926 has a quasi-periodic oscillation (QPO) similar to, but at a
much longer period than, the QPO seen in some Galactic Black Hole
Candidates. The power spectrum of the February data has several peaks, the
second highest of which is consistent with a monotonic decrease in the X-ray
period. The period change is then consistent with that expected from
two massive black holes spiralling together due to the emission of
gravitational radiation. This possibility is very unlikely but mentioned
because of its potential importance.
\end{abstract}


\section{Introduction}

We recently discovered a 16~hr X-ray periodicity in the Seyfert galaxy
IRAS~18325-5926 using data from ASCA (Iwasawa et al 1998). Nearly 9 cycles
of the oscillation were observed, with a total amplitude of about 15 per
cent. The active nucleus has similar properties to that of Seyfert 1
galaxies, except that the power-law continuum is slightly steeper than most
(photon index $\Gamma\sim 2.1$) and there is moderate absorption by a column
density of $\sim 10^{22}\pcmsq$. There is also a broad iron line in the
X-ray spectrum (Iwasawa et al 1996) which indicates the presence of an
accretion disk in the X-ray emission region, viewed at moderate inclination
(40--50 deg). The periodicity is plausible for Keplerian motion at 10--20
gravitational radii around a black hole of mass $2\times 10^8 - 2\times
10^7\Msun$. It also scales well with quasi-periodic oscillations (QPO) seen
from some Galactic Black Hole Candidates (BHC; Belloni et al 1997). The
cause of the oscillation in IRAS~18325-5926, or indeed in any BHC, is
unknown.

In order to determine whether the variation is exactly periodic or a QPO,
we observed IRAS~18325-5926 with the Rossi X-ray Timing Explorer in 1997
December 25--27 and 1998 February 21--23. We report here the light curves
and power spectra of those observations, which also show oscillations,
but of different periods. We conclude that the AGN has a clear QPO signal
and note that our results are consistent with the exciting possibility
that 2 black holes are rapidly spiralling together.

\section{Observations}
RXTE Proportional Counter Array (PCA) light curves were extracted from
stretches of data in which the number of Proportional Counter Units (PCUs)
on were consistent (e.g. light curves during observations in which only 3
PCUs were on were extracted separatedly from light curves during times when
all 5 PCUs were on).  We combined these to produce the final light curve by
scaling up to the level of 5 PCUs all those light curves with less than 5
PCUs on. Due to the potential urgency to report our results we have used
{\it realtime} data for the analysis of the February data; production data
are used for the December data.

Good time intervals were selected to exclude any Earth occultations or South
Atlantic Anomaly (SAA) passage and to ensure stable pointing. We generated
our background data using {\sc pcabackest v1.5} in order to estimate the
internal background caused by interactions between the radiation or
particles and the detector or spacecraft at the time of observation.  This
was done by matching the conditions of observations with those in various
model files.  The model files that we chose were constructed using the VLE
rate (one of the rates in PCA Standard 2 science array data that is defined
to be the rate of events which saturate the analog electronics) as the
tracer of the particle background.

Fig.~1 shows the background subtracted PCA light curve in the 4--10
 keV band for both observations in 1997 December and 1998 February. 
 

\begin{figure}
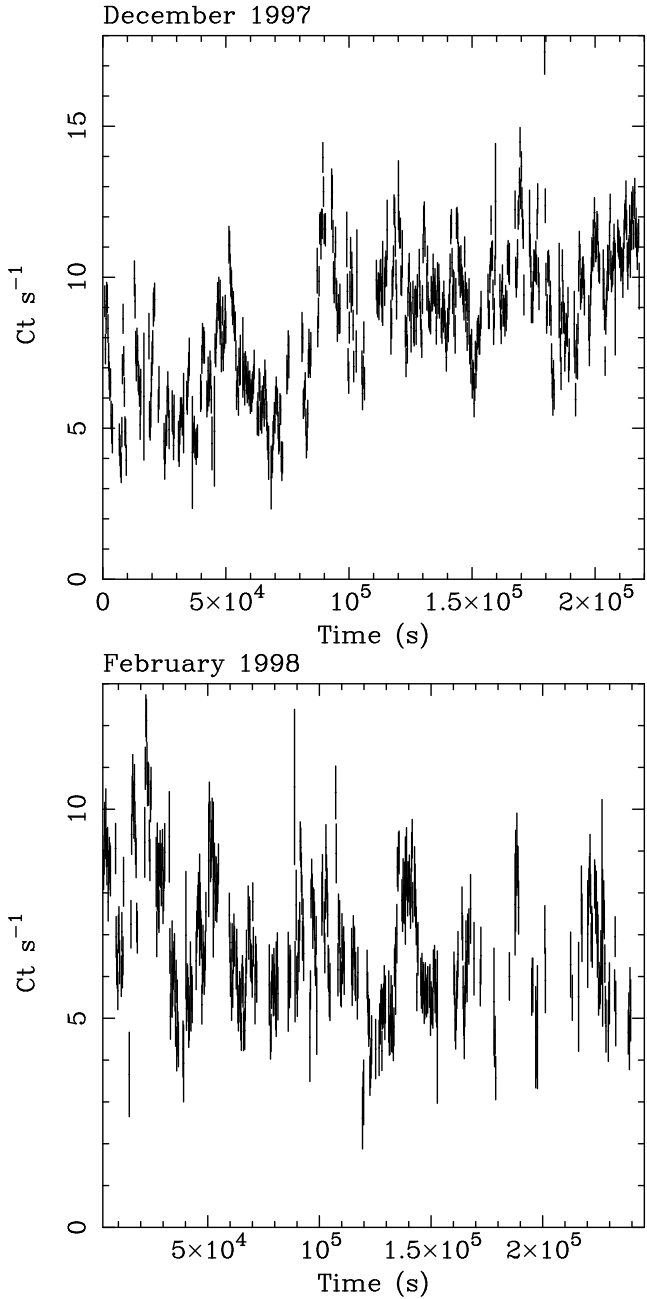

\centerline{\psfig{figure=dec_lc_new.ps,width=0.48\textwidth,angle=270}}
\centerline{\psfig{figure=feb_lc_new.ps,width=0.48\textwidth,angle=270}}
\caption{The RXTE light curve
of IRAS18325--5926 in the 4--10 keV band.}
\end{figure}

\section{Power spectra}

In the same manner as for the analysis of the ASCA data, we used the Lomb
algorithm (Lomb 1976; Press et al 1992) to determine the power spectra of
the RXTE lightcurves (Fig.~2). There is a clear peak in the first power
spectrum, at $2.52\times 10^{-5}\Hz$ (11.0~hr) in December 1997 and several
peaks in February 1998. The highest of these is at $2.48\times 10^{-5}\Hz$
(11.2~hr), very close to the December peak. The total amplitude of the
signal is strong, being about 50 per cent (see Fig.~3 and 4 for folded light
curves).

\begin{figure}
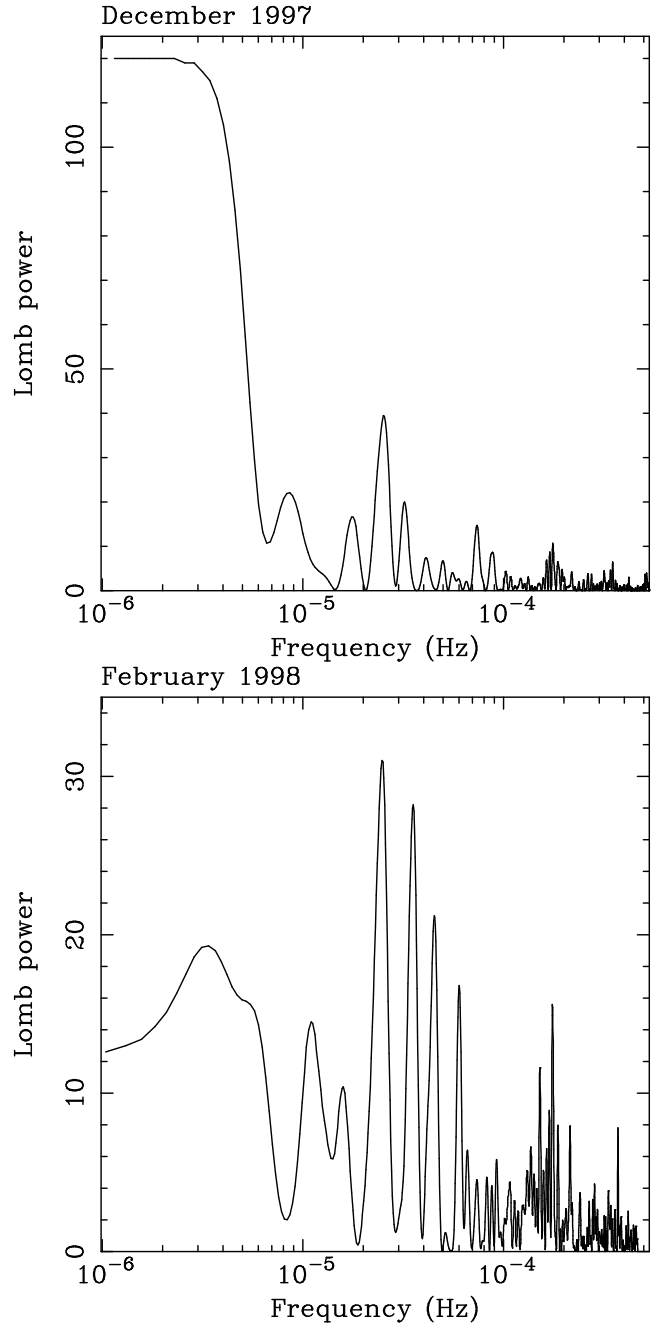

\centerline{\psfig{figure=dec_ps_new.ps,width=0.48\textwidth,angle=270}}
\centerline{\psfig{figure=feb_ps_new.ps,width=0.48\textwidth,angle=270}}
\caption{The power spectra derived from the RXTE light curve
of IRAS18325--5926 in the 4--10 keV band. }
\end{figure}

The arguments used in our previous work (Iwasawa et al 1998) for a
periodicity in each dataset being a real property of the Seyfert galaxy
apply here and are strengthened by the higher power of the signal.

We note that the power spike around $1.7\times 10^{-4}\Hz$ seen in the
spectrum of the February data is at the orbital period of the satellite, and
unlikely to be due to the source. We have examined the satellite
housekeeping and other available data and can find no explanation for the
main peaks in the power spectrum. The power spectrum of a 10 day long RXTE
observation of MCG--6-30-15 in August 1997 shows no distinct peaks (it shows
much red noise, as expected).

\begin{figure}
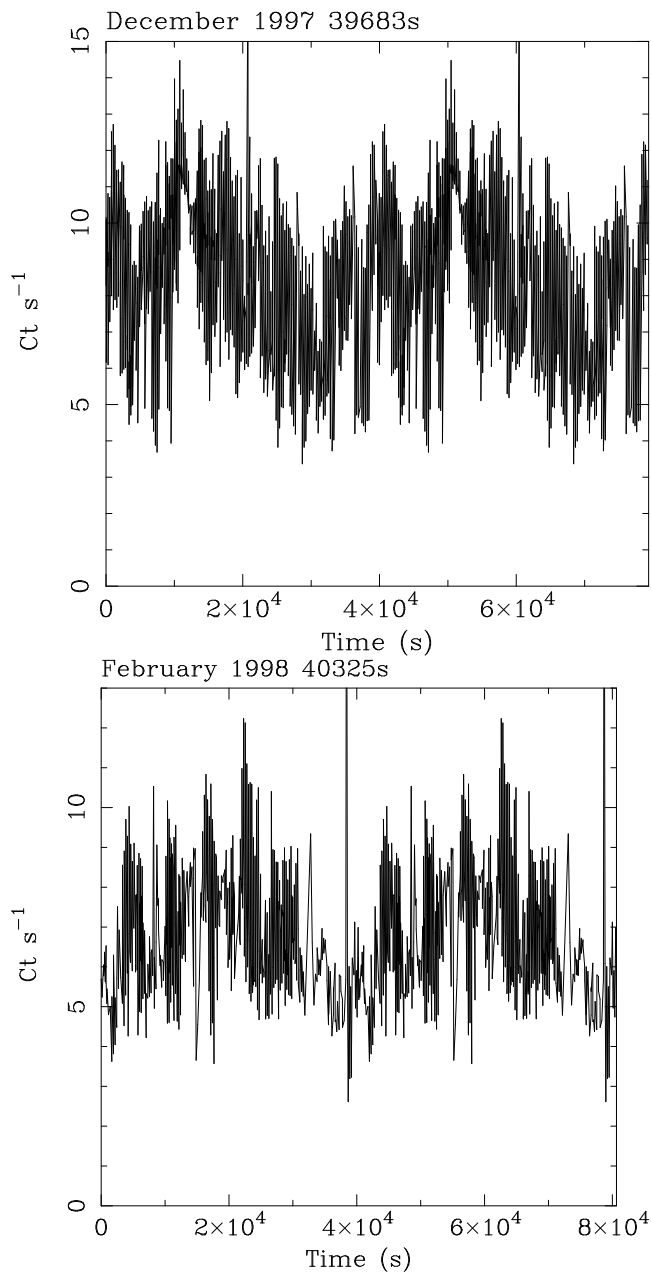

\centerline{\psfig{figure=fold_dec_new.ps,width=0.48\textwidth,angle=270}}
\centerline{\psfig{figure=fold_feb_new.ps,width=0.48\textwidth,angle=270}}
\caption{Lightcurves (2 periods of each are plotted) folded on the period of
the peak power spectrum frequency.}
\end{figure}

\begin{figure}
\centerline{\psfig{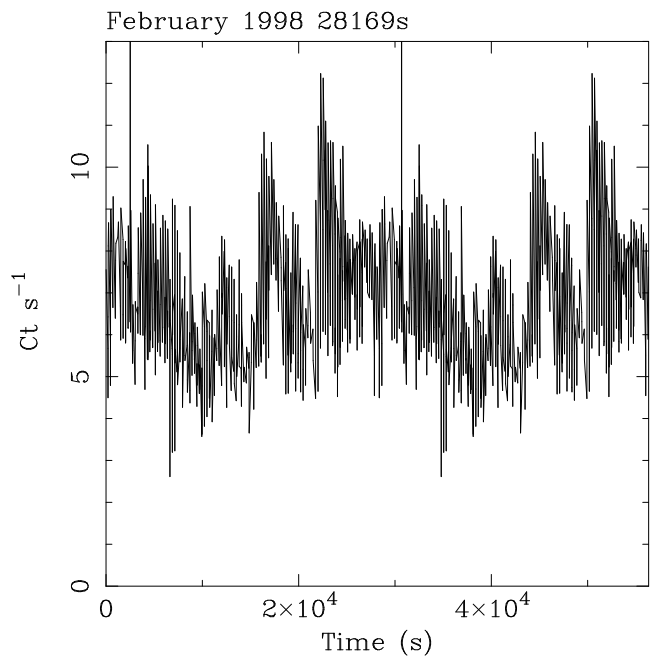}}
\caption{Lightcurves (2 periods of each are plotted) folded on the period of
the second peak of the February power spectrum.}
\end{figure}

\section{Discussion}

\begin{table}
\begin{center}
\caption{Mean flux and period for X-ray observations of IRAS~18325-5926.
The Ginga data are reported by Iwasawa et al (1995). The error bars indicate
the period change over which the power drops by a factor of two from the
peak. 1998a and b refer to the highest and next highest peaks in the
February power spectrum, respectively.}
\begin{tabular}{ccr}
Detector & Flux (4--10~keV) & Period  \\
 & $\ergpcmsqps$ & ks \\[5pt]
Ginga 1989  & $1.5\times 10^{-11}$ & $>30$ \\
ASCA\ 1994 & $5.6\times 10^{-12}$ & $>80$ \\
ASCA\ 1997 & $1.0\times 10^{-11}$ & $58.0^{+2.6}_{-1.8}$ \\
RXTE\ 1997 & $2.6\times10^{-11}$ & $38.7^{+4.2}_{-2.9}$  \\
RXTE\ 1998a & $2.1\times 10^{-11}$ & $40.3^{+4.5}_{-2.7}$ \\
RXTE\ 1998b & $2.1\times 10^{-11}$ & $28.2^{+1.9}_{-1.4}$ \\
\end{tabular}
\end{center}
\end{table}

The X-ray emission from IRAS~18325-5926 oscillates in a manner similar to
that seen in BHC QPO. Such a large clear oscillation from matter around a
black hole may suggest that we are detecting the fundamental frequency of
some space-filling corona above the disk. This may be possible if the corona
has a proton temperature close to the virial value and a lower electron
temperature (see e.g. Di Matteo, Blackman \& Fabian 1997). Of course this
must happen over a very restricted range of radii in order that a single
dominant oscillation is seen. Perhaps it corresponds to the radius where the
surface emission from the disk peaks (i.e. $\sim$7 Schwarzschild radii for a
non-spinning hole). The variation of period with flux (Fig.~5) is similar to
that seen in some QPO.

Otherwise, as mentioned by Iwasawa et al (1998), it could be due to the
Bardeen-Petterson effect if the angular momentum vectors of the black hole
and accreting matter are not aligned. A range of radii are then selected
over which the accretion disk tilts over to match the equatorial plane of
the black hole. (The disk is not actually precessing in this case; Markovic
\& Lamb 1998.) Reflection/obscuration by blobs of gas in the tilt zone might
then lead to observed flux variations, especially if the inclination is
fairly high (as the broad iron line may indicate; Iwasawa et al 1996), but
we consider it unlikely that 50 per cent variations can be obtained in this
way.

\begin{figure}
\centerline{\psfig{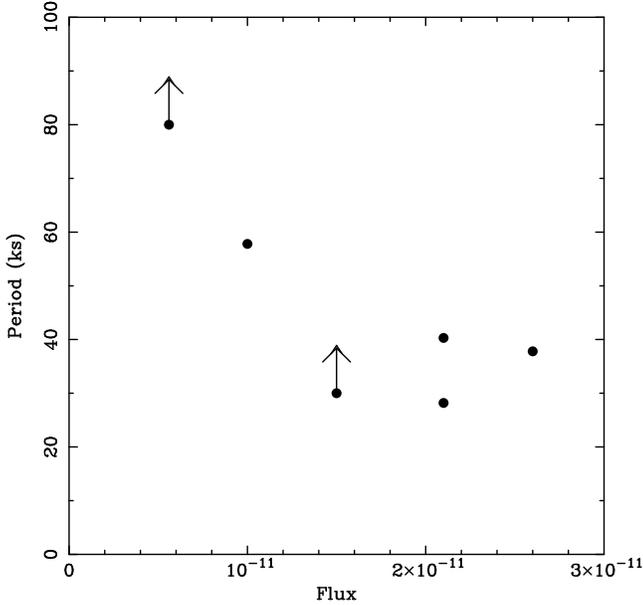}}
\caption{Variation of period with 4--10~keV flux. The two highest power
spectrum peaks are shown for February.}
\end{figure}

The data are consistent with a continuous decrease of period with time from
GINGA observations in 1989 to December 1997, and continuing if the second
highest peak in the February power spectrum is used. This raises the
possibility that the period is due to a massive object orbiting the black
hole (see Cunningham \& Bardeen 1973). The separation of the objects
requires that this second object is compact. The period decrease is then
explained as due to orbital energy loss by the emission of gravitational
waves (as for the binary pulsar). The rate of period change
$$\dot P ={{-96}\over5}{{G^{5/3}}\over c^5}{{M^{2/3}\mu(4\pi^2)^{4/3}}\over
P^{5/3}},$$
where $M=M_1+M_2$, the sum of the separate masses, and $\mu=M_1M_2/(M_1+M_2).$
This integrates to give $P\propto(t_0-t)^{3/8}$ where $t_0$ is the time when
the masses finally merge.

We have fitted this last relation to the data and find an acceptable fit
(Fig.~6). It suggests that the merger may occur in late April 1998. The
rate of spiral-in enables us to estimate that the product $M^{2/3}\mu\approx
1.5\times 10^{10}$ where the masses are in units of $\Msun$. This means that
for $M_1\sim 2\times 10^6\Msun - 10^8\Msun$, $M_2>10^5\Msun$. There is no
solution for $M_1<1.5\times 10^6\Msun$. Much of the energy from such a merger
would emerge as gravitational waves, but some is likely throughout the
electromagnetic spectrum.

We note that it is {\it a priori} unlikely to find an object close in time
to a spiral-in merger event. Possibilities that can enhance that probability
are if a) black holes are built out of many merger events where the basic
unit has a mass of only a few $10^5\Msun$ (say from dwarf galaxies) and b)
the inspiralling black hole switches on, or significantly enhances, an
otherwise quiescent active nucleus. If we make the extreme argument that
this process happens in all galaxies of space density
$3\times10^{-3}n_{-2.5}\Mpc^{-3}$, all of which have a central black hole of
mass $5\times10^7M_{7.5}\Msun$ growing by the addition of smaller black
holes of $10^5m_5\Msun$, then a merger will take place within $120\Mpc$ (the
distance to IRAS18325-5926) every $1000m_5 M_{7.5}^{-1} n_{-2.5}^{-1}\yr$. We
therefore see that such an event is not completely improbable but is
unlikely. The probability that we should find the signal of a merger in its
last year (it was the ASCA result of an observation in 1997 March which
alerted us to make the RXTE observations) is $\sim 10^{-3}$. Such mergers
would of course be more common at fainter flux levels ($\sim 1\yr^{-1}$
within 1 Gpc), and the prospects for space-based gravitational wave
astronomy (which will be sensitive to events in massive black holes) could
be very positive.

A lower mass black hole captured by a central black hole is likely initially
to have a highly eccentric orbit (Sigurdsson 1997). When the eccentricity
$e$ is high, the above estimates for $\dot P$ are dramatically increased
(Shapiro \& Teukolsky 1983), for example by a factor of 100--1000 when
$e\sim 0.8-0.9$, respectively. This allows $M_2$ to be much lower than
estimated above for a circular orbit (it can be as low as about $100\Msun$).
The overall temporal behaviour of the system is then more complex ($e$
decreases with time).  It may be difficult for a low mass black hole to
modulate the observed X-ray emission greatly; this problem is offset by the
enhanced probability of such an event occurring.

We stress that this last, exciting, interpretation involving an
in-spiralling black hole is unlikely and depends upon the identification of
the second highest power peak in the February 1998 data with the orbital
period of the second black hole. It does not explain the other peaks in the
power spectrum nor why some peaks expected in the lightcurve do not occur.
We discuss it here only because of its potential importance. Clearly further
observations are urgently required. Even if such observations fail to show a
consistently decreasing period, IRAS1832-5926 has a convincing,
high-amplitude, QPO, which requires explanation.

\begin{figure}
\centerline{\psfig{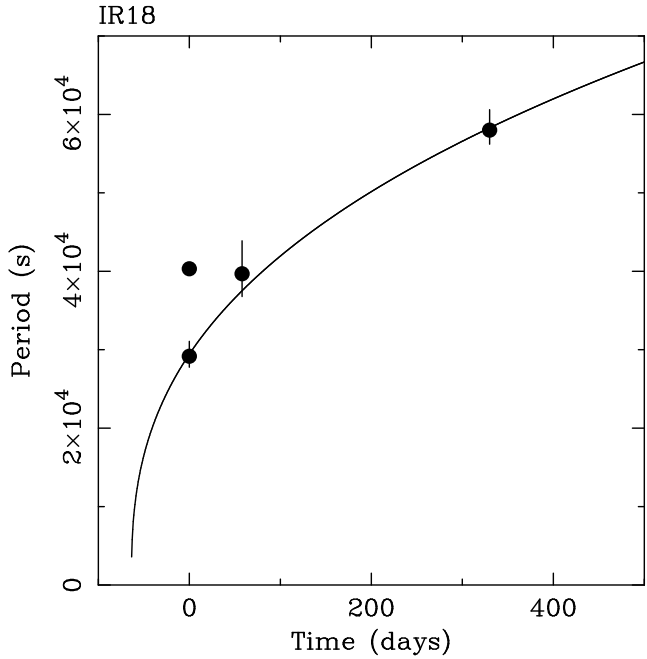}}
\caption{The peak period displayed against time in days, measured backwards
from the February 1998 observation. Error bars show where the peak power has
decreased by about a factor of 2. The line shows the best-fitting relation
$P=6203.(t+63.5)^{3/8}\s$. For the February data we plot both the period with
the peak power and the longer one which had slightly less power.}
\end{figure}

\section*{Acknowledgements}

We are grateful to Jean Swank for approving the February TOO observation and
to Steinn Sigurdsson for pointing out the potential importance of an
eccentric orbit. ACF and KI thank Royal Society and PPARC, respectively, for
support. JCL thanks the Isaac Newton Trust, the Overseas Research
Studentship programme (ORS) and the Cambridge Commonwealth Trust for
support. CSR acknowledges support from the NSF under grant AST9529175 and
NASA under grant NAG5-6337 and WNB support from NASA grant NAG5-6852.

\end{document}